\PassOptionsToPackage{unicode}{hyperref}
\PassOptionsToPackage{hyphens}{url}
\documentclass[twocolumn,10pt]{article}
\usepackage[letterpaper,margin=0.75in]{geometry}
\usepackage{mathptmx}
\usepackage{titlesec}
\titleformat*{\section}{\large\bfseries}
\titleformat*{\subsection}{\normalsize\bfseries}
\titleformat*{\subsubsection}{\normalsize\bfseries}
\titlespacing*{\section}{0pt}{2.0ex plus .4ex minus .2ex}{1.0ex plus .2ex}
\titlespacing*{\subsection}{0pt}{1.5ex plus .3ex minus .2ex}{0.7ex plus .1ex}
\titlespacing*{\subsubsection}{0pt}{1.2ex plus .3ex minus .2ex}{0.5ex plus .1ex}
\usepackage{xcolor}
\usepackage{amsmath,amssymb}
\usepackage{iftex}
\ifPDFTeX
  \usepackage[T1]{fontenc}
  \usepackage[utf8]{inputenc}
  \usepackage{textcomp} % provide euro and other symbols
\else % if luatex or xetex
  \usepackage{unicode-math} % this also loads fontspec
  \defaultfontfeatures{Scale=MatchLowercase}
  \defaultfontfeatures[\rmfamily]{Ligatures=TeX,Scale=1}
\fi
\usepackage{lmodern}
\ifPDFTeX\else
\fi
\IfFileExists{upquote.sty}{\usepackage{upquote}}{}
\IfFileExists{microtype.sty}{% use microtype if available
  \usepackage[]{microtype}
  \UseMicrotypeSet[protrusion]{basicmath} % disable protrusion for tt fonts
}{}
\makeatletter
\@ifundefined{KOMAClassName}{% if non-KOMA class
  \setlength{\parindent}{1em}
}{% if KOMA class
  \KOMAoptions{parskip=half}}
\makeatother
\usepackage{longtable,booktabs,array}
\usepackage{calc} % for calculating minipage widths
\usepackage{etoolbox}
\makeatletter
\patchcmd\longtable{\par}{\if@noskipsec\mbox{}\fi\par}{}{}
\makeatother
\IfFileExists{footnotehyper.sty}{\usepackage{footnotehyper}}{\usepackage{footnote}}
\makesavenoteenv{longtable}
\usepackage{graphicx}
\makeatletter
\newsavebox\pandoc@box
\newcommand*\pandocbounded[1]{% scales image to fit in text height/width
  \sbox\pandoc@box{#1}%
  \Gscale@div\@tempa{\textheight}{\dimexpr\ht\pandoc@box+\dp\pandoc@box\relax}%
  \Gscale@div\@tempb{\linewidth}{\wd\pandoc@box}%
  \ifdim\@tempb\p@<\@tempa\p@\let\@tempa\@tempb\fi% select the smaller of both
  \ifdim\@tempa\p@<\p@\scalebox{\@tempa}{\usebox\pandoc@box}%
  \else\usebox{\pandoc@box}%
  \fi%
}
\def\fps@figure{htbp}
\makeatother
\providecommand{\tightlist}{%
  \setlength{\itemsep}{0pt}\setlength{\parskip}{0pt}}
\usepackage{booktabs}
\usepackage{graphicx}
\usepackage{tikz}
\usetikzlibrary{arrows.meta, positioning, fit, backgrounds}
\usepackage{bookmark}
\IfFileExists{xurl.sty}{\usepackage{xurl}}{} % add URL line breaks if available
\makeatletter
\@ifundefined{xmpquote}{}{}
\makeatother
\definecolor{citeblue}{HTML}{0645AD}
\hypersetup{
  pdftitle={Adaptive Evaluation of Out-of-Band Defenses Against Prompt Injection in LLM Agents},
  pdfauthor={Praneeth Narisetty, Shiva Nagendra Babu Kore, Uday Kumar Reddy Kattamanchi, Jayaram Kumarapu},
  colorlinks=true,
  linkcolor=citeblue,
  citecolor=citeblue,
  urlcolor=black,
  pdfcreator={LaTeX via pandoc}}

\title{\huge\bfseries Adaptive Evaluation of Out-of-Band Defenses Against Prompt Injection in LLM Agents}
\author{Praneeth Narisetty, Shiva Nagendra Babu Kore,\\[2pt]
Uday Kumar Reddy Kattamanchi, Jayaram Kumarapu\\[3pt]
\href{https://launchsafe.com}{\textbf{LaunchSafe Research}}\\[2pt]
\{praneeth, shiva, uday, jayaram\}@launchsafe.com}
\date{}

\begin{document}
\maketitle
\begin{abstract}
Recent work (2024 to 2026) has converged on a strategy for defending
tool-using LLM agents against indirect prompt injection: rather than
training the model to refuse malicious instructions, enforce security
outside the model with a deterministic policy that mediates the agent's
actions. Systems such as CaMeL, FIDES, Progent, RTBAS, and FORGE realize
this with capabilities, information-flow labels, and reference monitors,
and several report near-elimination of attacks on the AgentDojo
benchmark. We make \textbf{two contributions}. First, we organize these
out-of-band defenses as instances of classical integrity protection
(Biba), reference monitoring, and least privilege, yielding a structured
comparison of what they do and do not cover. Second, we warn that every
one of them is validated only on \emph{static} benchmarks (a fixed set of
injection attempts), the same methodology that made in-band defenses look
strong until adaptive, defense-aware attacks broke twelve of them at over
90\% success; we specify the threat model and protocol an adaptive
evaluation requires. We then run that protocol as an \textbf{independent
reproduction and extension} of Progent's own adaptive-attack analysis, on
AgentDojo, with an open-weight agent (Qwen2.5-7B) self-hosted on a single
H200, a setting its authors did not test. Averaged over three runs,
\textbf{the defense held}: Progent cut mean attack success roughly
sixfold (25.8\% to 4.2\%), and a hand-crafted adaptive attack did not
raise it (2.6\%). This is one small-scale data point on a weak model with
a single black-box attack template; a stronger optimized (white-box GCG)
attack remains open. The result is \textbf{consistent with}, but does not
establish, the hypothesis that deterministic out-of-band enforcement is a
harder target for an adaptive attacker than in-band detection.
\end{abstract}

\vspace{0.5\baselineskip}
\noindent\textbf{Keywords:} prompt injection, indirect prompt injection, prompt injection defense, LLM agents, LLM agent security, AI agent security, agentic AI, autonomous agents, tool-using language models, tool use, function calling, out-of-band defense, reference monitor, privilege control, least privilege, Biba integrity, Saltzer--Schroeder, information-flow control, capabilities, taint tracking, adaptive evaluation, adaptive attacks, defense-aware attacks, adversarial robustness, red teaming, jailbreak, GCG, AgentDojo, ASB, Progent, CaMeL, FIDES, FORGE, RTBAS, Qwen2.5, LLM security, AI safety, AI security.

\subsection{1. Introduction}\label{introduction}

A production LLM agent is given tools, and through them, consequences:
it can read a database, call an internal API, send mail, move money. It
also reads text it does not control (a web page, a support ticket, a
PDF, the body of an email it was asked to summarize). When an attacker
plants instructions in that text and the agent follows them, the damage
is not a bad sentence but an action: a leaked record, a deleted row, a
transfer. This is prompt injection \hyperlink{cite:1}{\textbf{{[}1{]}}}, and in an agent with tools
it is an authorization problem, not a content problem.

The first generation of defenses treated it as content. Input
classifiers, guardrail models, jailbreak detectors, and adversarial
fine-tuning all try to recognize malicious text. By 2025 the evidence
against this approach was strong: adaptive attackers recover high
success rates against the detectors they were tested on \hyperlink{cite:34}{\textbf{{[}34{]}}},
\hyperlink{cite:36}{\textbf{{[}36{]}}}, and vendors building production agents now say plainly that
injection is unlikely to be fully solved at the model layer \hyperlink{cite:41}{\textbf{{[}41{]}}}.

The second generation gave up on the model and moved enforcement outside
it. If the model cannot be trusted to refuse, then wrap it in a layer
that decides what actions are permitted regardless of what the model was
talked into. CaMeL \hyperlink{cite:15}{\textbf{{[}15{]}}}, FIDES \hyperlink{cite:16}{\textbf{{[}16{]}}}, Progent \hyperlink{cite:17}{\textbf{{[}17{]}}}, RTBAS
\hyperlink{cite:20}{\textbf{{[}20{]}}}, Conseca \hyperlink{cite:18}{\textbf{{[}18{]}}}, and FORGE \hyperlink{cite:28}{\textbf{{[}28{]}}} are instances of this
move. They differ in mechanism, capabilities, taint labels, symbolic
privilege rules, isolation, but they share a structure that security
has used since the 1970s: a deterministic reference monitor enforcing a
policy at the point where an action takes effect. Several report strong
results. Progent cuts AgentDojo indirect-injection success from 39.9\%
to 1.0\% \hyperlink{cite:17}{\textbf{{[}17{]}}}; CaMeL solves 77\% of AgentDojo tasks under attack
with no successful injections in its threat model, against 84\%
undefended \hyperlink{cite:15}{\textbf{{[}15{]}}}.

This paper does two things. Sections 2--6 organize the second-generation
defenses through the classical primitives they instantiate, and compare
them on a common set of dimensions. This framing is not ours to claim, Zhang et al.~argue LLM agents should adopt the Saltzer--Schroeder
principles directly \hyperlink{cite:44}{\textbf{{[}44{]}}}, Bhattarai and Vu build an architecture on
reference monitors plus information-flow control \hyperlink{cite:27}{\textbf{{[}27{]}}}, and Shi et
al.~systematize the field through an authorization lens \hyperlink{cite:30}{\textbf{{[}30{]}}}. We use
the lens because it makes the comparison sharp, not because it is new.

The contribution is in Sections 7--10, and it is a problem the field has
not confronted. The second-generation defenses are evaluated the same
way the first generation was: against a fixed benchmark of attacks. That
method already failed once. Nasr et al.~took twelve published in-band
defenses, most reporting near-zero attack success, and recovered success
above 90\% for most of them with adaptive, defense-aware attacks
\hyperlink{cite:36}{\textbf{{[}36{]}}}. The action-level defenses have not faced this test. Their
reported numbers describe how they perform against attacks chosen before
the defense existed, not against an attacker who knows the defense
and optimizes against it. We argue this makes current confidence
premature, specify what an adaptive evaluation of an action-level
defense would require, and, new in this version (§11), execute
that evaluation against Progent as an independent reproduction and
extension of the adaptive-attack analysis its own authors report in
their Appendix E \hyperlink{cite:17}{\textbf{{[}17{]}}}.

\subsection{2. Prompt injection is an old vulnerability
class}\label{prompt-injection-is-an-old-vulnerability-class}

Prompt injection is the newest case of a failure that recurs wherever
one channel carries both control and data with no structural boundary
between them.

\begin{table}[h]
\centering\footnotesize
\caption{The same control/data confusion recurs across eras of computer security.}
\begin{tabular}{@{}
  >{\raggedright\arraybackslash}p{(\linewidth - 6\tabcolsep) * \real{0.2500}}
  >{\raggedright\arraybackslash}p{(\linewidth - 6\tabcolsep) * \real{0.2500}}
  >{\raggedright\arraybackslash}p{(\linewidth - 6\tabcolsep) * \real{0.2500}}
  >{\raggedright\arraybackslash}p{(\linewidth - 6\tabcolsep) * \real{0.2500}}@{}}
\toprule\noalign{}
\begin{minipage}[b]{\linewidth}\raggedright
Era
\end{minipage} & \begin{minipage}[b]{\linewidth}\raggedright
Vulnerability
\end{minipage} & \begin{minipage}[b]{\linewidth}\raggedright
What gets confused
\end{minipage} & \begin{minipage}[b]{\linewidth}\raggedright
What fixed it
\end{minipage} \\
\midrule\noalign{}
1990s & Buffer overflow, format string & Attacker data overruns into
control state & Bounds, W\^{}X, non-executable data \\
2000s (DB) & SQL injection & User input concatenated into a parsed query
& Parameterized queries \hyperlink{cite:12}{\textbf{{[}12{]}}} \\
2000s (web) & Cross-site scripting & User input rendered into parsed
markup & Output encoding; CSP \hyperlink{cite:13}{\textbf{{[}13{]}}}, \hyperlink{cite:14}{\textbf{{[}14{]}}} \\
2020s & Prompt injection & Untrusted text shares the token stream with
instructions & \emph{(open)} \\
\bottomrule\noalign{}
\end{tabular}
\end{table}

Schneier frames the root cause directly: mixing data with commands ``is
at the root of many of our computer security vulnerabilities'' \hyperlink{cite:3}{\textbf{{[}3{]}}}.
The StruQ authors place prompt injection in the family by name: it is
``yet another instance of this vulnerability pattern'' \hyperlink{cite:4}{\textbf{{[}4{]}}}. Willison
drew the SQL-injection parallel when he named the attack in 2022
\hyperlink{cite:2}{\textbf{{[}2{]}}}.

The cured members of the family teach one lesson: detection lost and
structure won. SQL injection was not solved by better recognition of
malicious queries but by parameterized queries, which fix the query's
structure before binding user input so that input cannot be parsed as
control \hyperlink{cite:12}{\textbf{{[}12{]}}}. XSS was contained by output encoding and a policy on
where executable content may come from \hyperlink{cite:13}{\textbf{{[}13{]}}}, \hyperlink{cite:14}{\textbf{{[}14{]}}}.

Prompt injection resists the same fix in one specific way. SQL has a
grammar, so a prepared statement can separate code from data because a
parser distinguishes them. Natural language has no such grammar, and the
model is trained to follow instructions wherever they appear. Zverev et
al.~measure this: current models do not maintain a usable separation
between instructions and data, and neither prompting nor fine-tuning
reliably induces one \hyperlink{cite:26}{\textbf{{[}26{]}}}. This is the structural fact the whole
field rests on: control/data separation cannot be enforced \emph{inside}
the model, so it must be enforced outside it. Figure 1 contrasts the two
postures.

\begin{figure*}[tb]
\centering\footnotesize
\begin{tikzpicture}[
  box/.style={draw, rounded corners=2pt, font=\footnotesize, align=center,
              minimum height=1.05cm, minimum width=1.9cm, inner sep=4pt},
  src/.style={box, minimum width=2.5cm, fill=black!4},
  proc/.style={box, fill=black!7},
  llmb/.style={box, fill=black!11},
  monb/.style={box, fill=black!22},
  ar/.style={-{Stealth[length=2.4mm]}, semithick},
  lbl/.style={font=\footnotesize\itshape, text=black!65}
]
% ---------- (a) In-band ----------
\node[font=\footnotesize\bfseries, anchor=west] at (0,3.3) {(a) In-band: control and data share one channel};
\node[src] (ti) at (1.35,2.4) {Trusted\\instructions};
\node[src] (ud) at (1.35,1.1) {Untrusted\\data};
\node[proc] (ts) at (4.7,1.75) {One token stream\\(no boundary)};
\node[llmb] (llm) at (7.6,1.75) {LLM};
\node[box] (act) at (10.0,1.75) {ACTION};
\draw[ar] (ti.east) -- (ts.west);
\draw[ar] (ud.east) -- (ts.west);
\draw[ar] (ts) -- (llm);
\draw[ar] (llm) -- (act);
\node[lbl, anchor=west] at (0.15,-0.1) {The model follows whichever instruction it reads, injected or not};
% ---------- (b) Out-of-band ----------
\node[font=\footnotesize\bfseries, anchor=west] at (0,-1.1) {(b) Out-of-band: a deterministic layer mediates the action};
\node[src] (tr) at (1.35,-1.95) {Trusted [HIGH]};
\node[src] (un) at (1.35,-3.25) {Untrusted [LOW]};
\node[llmb] (llm2) at (4.9,-2.6) {LLM\\(may be compromised)};
\node[proc] (prop) at (7.9,-2.6) {Proposed\\action};
\node[monb] (mon) at (10.4,-2.6) {POLICY\\MONITOR};
\node[box] (dec) at (12.9,-2.6) {Allow /\\deny};
\draw[ar] (tr.east) -- (llm2.west);
\draw[ar] (un.east) -- (llm2.west);
\draw[ar] (llm2) -- (prop);
\draw[ar] (prop) -- (mon);
\draw[ar] (mon) -- (dec);
\node[lbl, anchor=west] at (0.15,-4.3) {Provenance-tagged inputs; low-integrity data cannot authorize a high-integrity action};
\end{tikzpicture}
\caption{The same vulnerability, two postures. (a) In-band: control and data share one token stream. (b) Out-of-band: a deterministic monitor mediates the action.}
\end{figure*}

(The pre-digital precedent is in-band telephone signaling, where control
tones travelled on the voice channel and could be forged; the industry
moved signaling out of band. We note it as an analogy, not as a claim
about why SS7 was designed.)

\subsection{3. Threat model}\label{threat-model}

We use the threat model now standard in this literature \hyperlink{cite:15}{\textbf{{[}15{]}}},
\hyperlink{cite:16}{\textbf{{[}16{]}}}, \hyperlink{cite:37}{\textbf{{[}37{]}}}, and state it because disagreements about defense
effectiveness are often disagreements about the model.

\textbf{Assets.} Integrity of consequential actions (tool calls, writes,
sends, payments) and confidentiality of data the agent can reach (system
prompts, user records, retrieved private documents).

\textbf{Trust labels.} The system prompt is high-integrity. The primary
user's direct input is treated as high-integrity, with the caveat in
§8.3 that this can be wrong. Everything else (retrieved documents,
web pages, tool results, file contents, email, prior outputs of other
agents, persisted memory) is untrusted.

\textbf{Attacker.} Controls the content of at least one untrusted
channel, knows the system design including any deployed defense, and is
\emph{adaptive}: it can iterate, optimize adversarial strings, and craft
inputs against the specific monitor in place \hyperlink{cite:34}{\textbf{{[}34{]}}}, \hyperlink{cite:36}{\textbf{{[}36{]}}}. It does
not control the model weights or the enforcement layer.

\textbf{Direct vs.~indirect.} Direct injection arrives in the user's
turn; indirect injection arrives in content the agent consumes while
working \hyperlink{cite:1}{\textbf{{[}1{]}}}. Indirect injection is the harder case for tool-using
agents and our focus.

\textbf{Why actions are the point.} A chatbot that says something
harmful is a content failure; an agent that does something harmful is an
authorization failure. The defenses we study target actions. Section 8.1
returns to the harms this scope leaves out.

\subsection{4. In-band defenses provide no guarantee against adaptive
attackers}\label{in-band-defenses-provide-no-guarantee-against-adaptive-attackers}

By in-band defenses we mean those operating on or inside the model and
channel under attack: input/output classifiers, guardrail models,
instruction hierarchies, spotlighting, and adversarial fine-tuning
(StruQ, SecAlign, and successors). We report scores in the units the
sources use, attack success values (ASV) and false-negative rates
(FNR) on a 0--1 scale.

The instruction hierarchy leaks before any adaptive effort: on
GPT-4o-mini it admits existing combined attacks at ASV 0.68
(OpenPromptInjection) and 0.75 (MMLU-PI) \hyperlink{cite:34}{\textbf{{[}34{]}}}.

Fine-tuning defenses fall to adaptive optimization. Against adaptive
GCG, Jia et al.~drive StruQ to ASV \textasciitilde1.00 on MMLU-PI and
0.80--1.00 on OpenPromptInjection, and SecAlign to 0.88 on MMLU-PI (0.46
on OpenPromptInjection), far above the near-zero values originally
reported, and both lose utility under attack \hyperlink{cite:34}{\textbf{{[}34{]}}}. Pandya et
al.~break StruQ, SecAlign, and a successor with an architecture-aware
attack at up to 85--95\% success on unseen prompts \hyperlink{cite:35}{\textbf{{[}35{]}}}.

Detectors trade a flattering metric for a real blind spot. Attention
Tracker reaches AUC 1.00 on standard benchmarks but has FNR 0.69 on
MMLU-PI, and a separator-token-only adaptive attack pushes that to 1.00, a full bypass with trivial effort \hyperlink{cite:34}{\textbf{{[}34{]}}}. AUC on a static
benchmark overstates resilience.

The clearest result is Nasr et al.: twelve published defenses, most
originally reporting near-zero attack success, broken at above 90\% for
most under adaptive attack, spotlighting and sandwiching above 95\%,
PromptGuard and Model Armor above 90\%, MetaSecAlign at 96\%, with
known-answer detectors and others also falling; human red-teaming
reached 100\% \hyperlink{cite:36}{\textbf{{[}36{]}}}.

A guardrail model used to screen for injections is itself a model and is
itself injectable, as OWASP's guidance notes \hyperlink{cite:39}{\textbf{{[}39{]}}}, which follows
directly from §2.

Industry has conceded the model-layer limit. OpenAI states injection is
``unlikely to ever be fully `solved'\,'' \hyperlink{cite:41}{\textbf{{[}41{]}}}. Anthropic's strongest
reported numbers for Claude Opus 4.5 are real but not guarantees: against
indirect prompt injection in agentic coding environments (Gray Swan's Shade
adaptive red-teaming), attack success rises from 4.7\% at
one attempt to 33.6\% at ten and 63.0\% at one hundred \hyperlink{cite:43}{\textbf{{[}43{]}}}. A
defense whose failure probability climbs toward certainty under
repetition reduces incidents; it does not bound them.

Two clarifications matter for what follows. First, ``no guarantee'' is
the defensible claim, and it is a \emph{structural} one: §2 shows the
model has no reliable instruction/data boundary, so no amount of in-band
training or filtering can promise to refuse. It is not the stronger
claim that in-band defenses always fail, some hold non-adaptive
attacks to low success rates, and they have value as one layer of
defense-in-depth. Second, and this is the thread of the paper: the
reason adaptive attacks are decisive here is that the in-band defenses
were \emph{believed} effective on the strength of static benchmarks, and
the belief did not survive contact with an attacker who optimized
against them.

\subsection{5. The classical lens}\label{the-classical-lens}

Enforcement that the model cannot provide must come from a deterministic
mechanism around it. Security named that mechanism long ago, and reading
the modern defenses through it makes them comparable. We present this as
a lens, not a discovery; prior work uses the same vocabulary \hyperlink{cite:27}{\textbf{{[}27{]}}},
\hyperlink{cite:30}{\textbf{{[}30{]}}}, \hyperlink{cite:44}{\textbf{{[}44{]}}}.

\textbf{Biba integrity.} Biba's model protects against improper
modification with two rules: a subject may not read data below its
integrity level without being lowered to that level (Simple Integrity,
the low-water-mark reading), and a subject may not write above its level
(the Star Integrity property, ``no write up'') \hyperlink{cite:5}{\textbf{{[}5{]}}}. Applied to an
agent: when the model reads untrusted data its effective integrity
drops, and a dropped subject may not authorize a high-integrity action.
Both rules are needed, the first to capture the contamination, the
second to block the action. Figure 2 shows the invariant.

\begin{figure*}[tb]
\centering\footnotesize
\begin{tikzpicture}[
  band/.style={draw, rounded corners=2pt, minimum width=12cm, minimum height=10mm, align=center, font=\footnotesize, inner sep=5pt},
  hi/.style={band, fill=black!5},
  lo/.style={band, fill=black!15},
  ar/.style={-{Stealth[length=2.6mm]}, semithick},
  note/.style={font=\scriptsize, align=center, text=black!80}
]
\node[hi] (high) at (0,1.45) {\textbf{HIGH}\quad System prompt; authority to call a consequential tool};
\node[lo] (low)  at (0,-1.45) {\textbf{LOW}\quad Web pages, emails, tool results, files (attacker-influenceable)};
% read-down: low-water-mark
\draw[ar] (-5,0.9) -- (-5,-0.9);
\node[note, anchor=west] at (-4.6,0) {Reading low data lowers\\the subject (low-water-mark)};
% no write up: blocked
\draw[ar, red!75!black] (5,-0.9) -- (5,0.9);
\draw[red!75!black, line width=1.1pt] (4.82,-0.18) -- (5.18,0.18);
\draw[red!75!black, line width=1.1pt] (4.82,0.18) -- (5.18,-0.18);
\node[note, anchor=east, text=black] at (4.6,0) {\textbf{No write up}: cannot\\authorize a high action};
\end{tikzpicture}
\caption{The Biba integrity invariant on an agent's action surface. Reading low-integrity data lowers the subject (Simple Integrity / low-water-mark); a lowered subject may not authorize a high-integrity action (no write up). The only sanctioned upward path is endorsement from a trusted channel.}
\end{figure*}
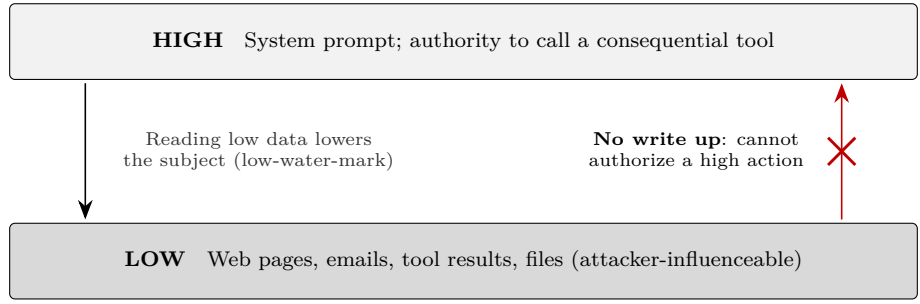

\textbf{Reference monitor.} Anderson's reference monitor validates every
access against the policy, and must be always invoked (complete
mediation), tamperproof, and small enough to verify \hyperlink{cite:6}{\textbf{{[}6{]}}}. Every
credible action-level defense is a reference monitor at the tool
boundary; the three requirements give a checklist, and a vocabulary for
failures (a side channel is incomplete mediation; an LLM that authors
the policy strains verifiability).

\textbf{Saltzer--Schroeder.} The 1975 principles map onto agent
security: complete mediation, least privilege, fail-safe defaults, and
economy of mechanism \hyperlink{cite:7}{\textbf{{[}7{]}}}. Progent's per-tool-call checking enforcing
least privilege \hyperlink{cite:17}{\textbf{{[}17{]}}} and the preference for a small deterministic
policy engine over a model-based judge are direct applications. Zhang et
al.~make the same mapping the center of their proposal \hyperlink{cite:44}{\textbf{{[}44{]}}}.

\textbf{Capabilities and information flow.} A capability is an
unforgeable token carrying both the designation of an object and the
rights to it, which removes ambient authority \hyperlink{cite:8}{\textbf{{[}8{]}}}, \hyperlink{cite:9}{\textbf{{[}9{]}}}, 
CaMeL's ``capability'' tags a value's provenance and permitted readers
and is checked at tool-call sinks \hyperlink{cite:15}{\textbf{{[}15{]}}}. Information-flow control
labels data and propagates the labels under a lattice order \hyperlink{cite:10}{\textbf{{[}10{]}}},
with the classic difficulty being implicit flows through control
decisions \hyperlink{cite:11}{\textbf{{[}11{]}}}; FIDES's taint labels \hyperlink{cite:16}{\textbf{{[}16{]}}} and the IFC framing of
Wu et al.~\hyperlink{cite:21}{\textbf{{[}21{]}}} instantiate it, and the implicit-flow problem is
exactly the side channel CaMeL reports against itself.

Read this way, the defenses line up: Dual-LLM \hyperlink{cite:25}{\textbf{{[}25{]}}} separates
subjects by integrity; CaMeL \hyperlink{cite:15}{\textbf{{[}15{]}}} adds capabilities and a
deterministic interpreter; FIDES \hyperlink{cite:16}{\textbf{{[}16{]}}} is taint tracking with
integrity and confidentiality labels; Progent \hyperlink{cite:17}{\textbf{{[}17{]}}} is least
privilege via symbolic per-call rules; RTBAS \hyperlink{cite:20}{\textbf{{[}20{]}}} gates tool calls
on flow labels (and we read it as enforcing the same integrity
invariant, though it does not invoke Biba by name); FORGE \hyperlink{cite:28}{\textbf{{[}28{]}}} is a
reference monitor over Datalog policies. The lens organizes them; it
does not, by itself, tell us which ones survive a real attacker.

\subsection{6. A systematization of out-of-band
defenses}\label{a-systematization-of-out-of-band-defenses}

We compare the principal defenses on eight dimensions: (D1) enforcement
primitive; (D2) is the gate deterministic or an LLM in the loop; (D3)
where the monitor sits; (D4) integrity (action) coverage; (D5)
confidentiality coverage; (D6) implicit-flow / side-channel handling;
(D7) reported cost; (D8) can it retrofit onto an unmodified agent. These
dimensions follow from the reference-monitor and IFC framing of §5; we
do not claim they are the only possible axes, and the classification of
each row reflects the sources' own descriptions.

\begin{table*}[t]
\centering\scriptsize\setlength{\tabcolsep}{3pt}
\caption{Systematization of out-of-band defenses across eight dimensions (D1--D8).}
\begin{tabular}{@{}
  >{\raggedright\arraybackslash}p{(\linewidth - 16\tabcolsep) * \real{0.1111}}
  >{\raggedright\arraybackslash}p{(\linewidth - 16\tabcolsep) * \real{0.1111}}
  >{\raggedright\arraybackslash}p{(\linewidth - 16\tabcolsep) * \real{0.1111}}
  >{\raggedright\arraybackslash}p{(\linewidth - 16\tabcolsep) * \real{0.1111}}
  >{\raggedright\arraybackslash}p{(\linewidth - 16\tabcolsep) * \real{0.1111}}
  >{\raggedright\arraybackslash}p{(\linewidth - 16\tabcolsep) * \real{0.1111}}
  >{\raggedright\arraybackslash}p{(\linewidth - 16\tabcolsep) * \real{0.1111}}
  >{\raggedright\arraybackslash}p{(\linewidth - 16\tabcolsep) * \real{0.1111}}
  >{\raggedright\arraybackslash}p{(\linewidth - 16\tabcolsep) * \real{0.1111}}@{}}
\toprule\noalign{}
\begin{minipage}[b]{\linewidth}\raggedright
System
\end{minipage} & \begin{minipage}[b]{\linewidth}\raggedright
D1
\end{minipage} & \begin{minipage}[b]{\linewidth}\raggedright
D2
\end{minipage} & \begin{minipage}[b]{\linewidth}\raggedright
D3
\end{minipage} & \begin{minipage}[b]{\linewidth}\raggedright
D4
\end{minipage} & \begin{minipage}[b]{\linewidth}\raggedright
D5
\end{minipage} & \begin{minipage}[b]{\linewidth}\raggedright
D6
\end{minipage} & \begin{minipage}[b]{\linewidth}\raggedright
D7 cost (reported)
\end{minipage} & \begin{minipage}[b]{\linewidth}\raggedright
D8 retrofit
\end{minipage} \\
\midrule\noalign{}
Dual-LLM \hyperlink{cite:25}{\textbf{{[}25{]}}} & subject separation & deterministic controller &
orchestrator & yes & partial & not addressed & design only & no
(rebuild) \\
CaMeL \hyperlink{cite:15}{\textbf{{[}15{]}}} & capabilities + CFI & deterministic interpreter & custom
interpreter & yes & yes & side channels shown by authors & 77\% vs 84\%
tasks (o3, under attack); 2.82$\times$ in / 2.73$\times$ out tokens & no (rewrite
agent) \\
FIDES \hyperlink{cite:16}{\textbf{{[}16{]}}} & taint labels (C+I) & deterministic sink check & planner
& yes & yes & improves on prior IFC; trade-offs stated & no headline
figure & no (adopt planner) \\
Progent \hyperlink{cite:17}{\textbf{{[}17{]}}} & symbolic privilege rules & deterministic check;
LLM-authored policy & tool-call layer & yes & partial & n/a & AgentDojo
39.9\%$\rightarrow$1.0\%; ASB 70.3\%$\rightarrow$3.9\% & \textbf{yes (proxy mode)} \\
Conseca \hyperlink{cite:18}{\textbf{{[}18{]}}} & JIT policy from trusted context & deterministic &
planner / executor & yes & partial & adversarial-planning risk & not
reported & no (planner split) \\
RTBAS \hyperlink{cite:20}{\textbf{{[}20{]}}} & IFC + screeners & mixed (LM-judge screener) & tool-call
gating & yes & yes & partly heuristic & \textasciitilde2\% utility loss
& partial (LLM in loop) \\
FORGE \hyperlink{cite:28}{\textbf{{[}28{]}}} & Datalog reference monitor & deterministic & action
boundary & yes & partial & policy-dependent & not reported & \textbf{yes
(no agent change)} \\
\bottomrule\noalign{}
\end{tabular}
\end{table*}

Three points. The gates are deterministic where it counts (D2): the
field learned that the gate must not be a model. Confidentiality and
implicit flows are the weak columns (D5, D6), most systems gate
actions well and handle exfiltration and side channels poorly, a point
§8 develops. And contrary to a common impression, retrofit is not the
open problem (D8): Progent's proxy mode applies ``without altering the
agent's internal implementation'' \hyperlink{cite:17}{\textbf{{[}17{]}}}, and FORGE enforces policies
``without modification to the underlying agents'' \hyperlink{cite:28}{\textbf{{[}28{]}}}. Deployment
onto existing agents is, at least in prototype, already solved.

(D7 mixes units, task percentages, token multipliers, utility loss, because the sources do; it should be read as ``what each paper
reports,'' not a normalized comparison.)

\subsection{7. The gap is evaluation, not
deployment}\label{the-gap-is-evaluation-not-deployment}

If retrofit is solved and the mechanisms are sound, what is missing? The
answer is in column D7 and in how every cell in the table was produced.
Most of each defense's headline security is reported against a
\emph{static} benchmark: AgentDojo's fixed injection set \hyperlink{cite:37}{\textbf{{[}37{]}}}, ASB,
or similar. The attacks were fixed before the defense existed. Adaptive
evaluation has begun in places, Progent's authors construct adaptive
attacks against their own policy-update LLM in Appendix E \hyperlink{cite:17}{\textbf{{[}17{]}}}, and
AgentDyn \hyperlink{cite:45}{\textbf{{[}45{]}}} re-evaluates Progent and CaMeL on harder dynamic tasks, but the field still lacks a standardized, independent, defense-aware
adaptive protocol applied across these systems, attack families
(black-box \emph{and} white-box), and models.

This is the precise methodology that failed for in-band defenses. StruQ,
SecAlign, PromptGuard, Spotlighting, and the rest reported near-zero
attack success on static benchmarks. Then Jia et al.~and Nasr et al.~let
the attacker move second, and the numbers inverted, twelve defenses
above 90\% \hyperlink{cite:34}{\textbf{{[}34{]}}}, \hyperlink{cite:36}{\textbf{{[}36{]}}}. The static benchmark did not measure what
it was taken to measure. It measured resistance to a known attack set,
and was read as resistance to attackers.

We have no result showing the action-level defenses would fall the same
way. Their gates are deterministic, which is a harder target than a
classifier, and an adaptive attacker faces a different problem: not
``evade a detector'' but ``drive a consequential action while respecting
the policy.'' That may be genuinely hard. But it is untested. The
defenses' authors are often careful about this, CaMeL demonstrates
side channels against itself \hyperlink{cite:15}{\textbf{{[}15{]}}}, Progent constructs adaptive
attacks on its own policy LLM \hyperlink{cite:17}{\textbf{{[}17{]}}}, and FORGE and Conseca name
adversarial-planning risks, yet the systematic, independent,
multi-attack adaptive evaluation that the in-band defenses received has
not been assembled for this class. Until that happens, the field is
largely in the position it was in for in-band defenses just before they
broke: confident, on the basis of an evaluation method with a
demonstrated blind spot.

So the open question is not ``can we deploy a reference monitor onto an
existing agent'' (yes) but ``does the reference monitor hold when the
attacker optimizes against it'' (unknown). That is the evaluation this
area lacks, and it is where effort should go.

\subsection{8. What remains even if the defenses
hold}\label{what-remains-even-if-the-defenses-hold}

Set aside adaptive robustness for a moment. The action-mediation
approach has limits that hold by construction, which a defender adopting
any system in §6 inherits. We lead with the two that are least
discussed.

\subsubsection{8.1 Adaptive evaluation is missing (and so is the
methodology for
it)}\label{adaptive-evaluation-is-missing-and-so-is-the-methodology-for-it}

This is §7 restated as an agenda item. While individual papers probe
adaptive robustness (Progent's Appendix E \hyperlink{cite:17}{\textbf{{[}17{]}}}, AgentDyn \hyperlink{cite:45}{\textbf{{[}45{]}}}),
there is still no \emph{standardized}, independent, defense-aware
evaluation applied uniformly across action-level defenses, and no agreed
threat model for what ``the attacker optimizes against the monitor''
means when the monitor is deterministic and the attacker's lever is the
model's reasoning. Wang et al.~and Shi et al.~catalog defense gaps
\hyperlink{cite:30}{\textbf{{[}30{]}}}, \hyperlink{cite:31}{\textbf{{[}31{]}}}, but neither runs this evaluation. Section 10
specifies what it would take.

\subsubsection{8.2 Provenance assignment is the trusted base, and it is
under-specified}\label{provenance-assignment-is-the-trusted-base-and-it-is-under-specified}

Every labeling scheme rests on an oracle that assigns initial
provenance, and the universal simplification is that the primary user's
input is trusted. Willison names the failure: if a user can be tricked
into pasting untrusted content, the labels are wrong at the source
\hyperlink{cite:25}{\textbf{{[}25{]}}}. FIDES's deployed form inherits the dual risk, a forgotten
label defaults to trusted \hyperlink{cite:16}{\textbf{{[}16{]}}}. Provenance is the trusted computing
base of the whole approach, and assigning it correctly at real system
boundaries (which header, which span, which upstream agent) is where
deployments will fail. It is under-specified in every system we
surveyed.

\subsubsection{8.3 In-the-loop tasks}\label{in-the-loop-tasks}

The clean defenses work by keeping untrusted data out of the trusted
plan. But many real tasks require the model to act on untrusted content, ``read this email and, if it's a meeting request, add it to my
calendar.'' The untrusted data must drive a consequential action. The
current answer is to ask the user to endorse the action (RTBAS, CaMeL)
\hyperlink{cite:20}{\textbf{{[}20{]}}}, \hyperlink{cite:15}{\textbf{{[}15{]}}}, which both reintroduces a human judgment an attacker
can target and produces approval fatigue. A principled account of which
in-the-loop tasks are securable does not exist, and is the deepest open
problem here.

\subsubsection{8.4 Text-to-text harms}\label{text-to-text-harms}

Action mediation protects actions. It does nothing about injections
whose payload is text, a poisoned document that yields a misleading
summary, or an injected instruction to the user. CaMeL states this as an
explicit non-goal \hyperlink{cite:15}{\textbf{{[}15{]}}}. As agents produce more text that people act
on, this undefended channel grows.

\subsubsection{8.5 Implicit flows and side
channels}\label{implicit-flows-and-side-channels}

CaMeL demonstrates two working side channels against itself (an
indirect-dependency leak and an attacker-triggered exception leak) and
discusses a timing channel its implementation mitigates, concluding it
``is vulnerable to side-channel attacks'' \hyperlink{cite:15}{\textbf{{[}15{]}}}. This is the
classical implicit-flow problem \hyperlink{cite:11}{\textbf{{[}11{]}}}. Taint systems that track only
explicit data dependence miss these by construction, and sound handling
pulls against the small-and-verifiable ideal of a reference monitor.

\subsection{9. Implications for a retrofit
defense}\label{implications-for-a-retrofit-defense}

Because retrofit is already feasible (§6) and adaptive robustness is the
real unknown (§7), the useful design question is narrow: what would a
retrofit monitor need to do that current ones do not? One under-explored
direction is provenance-awareness. Progent's proxy checks tool-call
names and arguments \hyperlink{cite:17}{\textbf{{[}17{]}}}; it does not track the provenance of the
data those arguments derive from. A monitor that consulted transitive
provenance could enforce the Biba invariant directly rather than
approximating it with argument patterns. The obstacle is real and worth
stating: a layer that sees only tool I/O cannot observe how the model
laundered low-integrity text into a high-integrity argument inside its
hidden reasoning (§8.5). Whether provenance-aware retrofit is achievable
without instrumenting the model is an open question, and we do not claim
to have answered it. We flag it as the design problem that follows from
the systematization, not as a finished contribution.

\subsection{10. A protocol for adaptive
evaluation}\label{a-protocol-for-adaptive-evaluation}

The paper's concrete deliverable is a specification of the evaluation §7
calls for, so that it can be run by us or others. §11 reports our first
execution of it; this section states the protocol in general form.

\textbf{Targets.} Open or reproducible action-level defenses, Progent
(open source), CaMeL, and any with available implementations, on
AgentDojo \hyperlink{cite:37}{\textbf{{[}37{]}}} (97 tasks, 629 security cases) and ASB.

\textbf{Attacker.} Defense-aware and adaptive, following Jia et al.~and
Nasr et al.~\hyperlink{cite:34}{\textbf{{[}34{]}}}, \hyperlink{cite:36}{\textbf{{[}36{]}}}: adaptive string optimization against the
deployed monitor; provenance-spoofing where labels are
attacker-influenceable; endorsement-targeting social engineering for
in-the-loop tasks; and side-channel probes of deterministic policies.
The attacker knows the policy.

\textbf{Metrics.} Action-level ASV/FNR on a 0--1 scale; utility
retention against an undefended baseline; endorsement-request rate (a
usability and §8.3-exposure proxy); and monitor overhead.

\textbf{What would the result mean.} If deterministic action monitors
hold near-zero ASV under a \emph{strong, optimized} adaptive attack,
that is evidence that the second-generation defenses are meaningfully
stronger than the first, and it would help justify the field's current
confidence. If they fall the way the in-band defenses did, the field is
repeating its mistake and the reported numbers are not measuring
security. Either outcome is worth knowing, and neither is known today.

\subsection{11. Empirical evaluation: reproducing and extending
Progent's adaptive-attack setting on an open-weight
model}\label{empirical-evaluation-reproducing-and-extending-progents-adaptive-attack-setting-on-an-open-weight-model}

We ran the §10 protocol against Progent, the open-source, deployable
out-of-band defense \hyperlink{cite:17}{\textbf{{[}17{]}}}, as an \textbf{independent reproduction
and extension} of the adaptive-attack analysis its authors report in
their Appendix E. Progent's authors construct adaptive attacks against
their own LLM policy-update mechanism (under a vulnerable-user
auto-approve simulation) and report that attack success stays low (they
report roughly 0.5--4.2\% across their constructions, with utility
retained). They ran this on GPT-4o; we ask whether the same holds on a
weak, self-hosted open-weight agent, and we add repeated runs. The
central question is unchanged: \emph{does the deterministic gate hold
when the attacker adapts to it?}, but our contribution is a
reproduction on a different model with variance reported, not a
first-of-its-kind test.

\subsubsection{11.1 Who and how}\label{who-and-how}

The evaluation was conducted by LaunchSafe Research on a single NVIDIA
H200 GPU. The full stack is open-weight and self-hosted (no external
model API):

\begin{itemize}
\tightlist
\item
  \textbf{Agent model.} Qwen2.5-7B-Instruct, served with vLLM (32K
  context, native tool-calling).
\item
  \textbf{Benchmark.} AgentDojo \hyperlink{cite:37}{\textbf{{[}37{]}}}, the standard tool-using-agent
  injection benchmark. Suites: \textbf{banking, slack, workspace}
  (travel excluded; see §12). We use a fixed subsample, the first 8
  user tasks per suite $\times$ all injection tasks, held identical across
  every condition so the comparison is valid.
\item
  \textbf{Defense.} Progent (\texttt{sunblaze-ucb/\allowbreak progent}), default
  configuration (policy updates permitted, not narrow-only). Progent's
  policy-authoring LLM was set to the same local Qwen2.5-7B (the authors
  used GPT-4o; see the policy-model substitution note in §12).
\item
  \textbf{Conditions, per suite.} (A) undefended + standard attack; (B)
  Progent + standard attack; (C) Progent + our adaptive attack.
\item
  \textbf{Repeats.} The entire study was run \textbf{three times}
  (identical configuration, agent temperature 0). We report mean $\pm$
  standard deviation across the three runs, which both averages out
  residual nondeterminism and tells us how reproducible the result is.
\item
  \textbf{Standard attack.} AgentDojo's
  \texttt{important\_\allowbreak{}instructions}.
\item
  \textbf{Adaptive attack (ours).} A defense-aware injection that,
  beyond instructing the agent, disguises the malicious action as a
  \emph{benign, pre-authorized, necessary} step, crafted to make
  Progent's policy-\emph{update} model (which widens the allowlist based
  on the ``benign content'' of tool results) authorize the malicious
  tool call. This targets the one model-based component of an otherwise
  deterministic defense, i.e., the ``LLM that authors the policy strains
  verifiability'' concern from §5.
\item
  \textbf{Metric.} AgentDojo's per-(user task, injection task)
  \texttt{security} flag = \textbf{attack success} (the injection goal
  was executed). We verified this orientation directly: it \emph{falls}
  when the defense is enabled (matching Progent's reported direction)
  and is bounded by the agent's ability to perform the injection goal at
  all.
\end{itemize}

\subsubsection{11.2 Results}\label{results}

\begin{table}[h]
\centering\scriptsize\setlength{\tabcolsep}{3pt}
\caption{Attack success rate (ASR) by application and condition (mean $\pm$ std over 3 runs; lower is safer).}
\begin{tabular}{@{}llll@{}}
\toprule\noalign{}
Application & Undefended & Progent (standard) & Progent (adaptive) \\
\midrule\noalign{}
banking & 19.0\% $\pm$ 0.7 & 1.9\% $\pm$ 0.7 & 3.7\% $\pm$ 1.3 \\
slack & 58.3\% $\pm$ 6.2 & 10.8\% $\pm$ 1.2 & 4.2\% $\pm$ 1.2 \\
workspace & 0.0\% $\pm$ 0.0 & 0.0\% $\pm$ 0.0 & 0.0\% $\pm$ 0.0 \\
\textbf{mean} & \textbf{25.8\% $\pm$ 2.1} & \textbf{4.2\% $\pm$ 0.3} &
\textbf{2.6\% $\pm$ 0.04} \\
\bottomrule\noalign{}
\end{tabular}
\end{table}

\begin{table}[h]
\centering\scriptsize\setlength{\tabcolsep}{3pt}
\caption{Task utility under attack by application and condition (mean over 3 runs; higher is better).}
\begin{tabular}{@{}llll@{}}
\toprule\noalign{}
Application & Undefended & Progent (standard) & Progent (adaptive) \\
\midrule\noalign{}
banking & 31.9\% & 31.9\% & 31.0\% \\
slack & 47.5\% & 20.8\% & 19.2\% \\
workspace & 56.3\% & 28.5\% & 27.1\% \\
\textbf{mean} & \textbf{45.2\%} & \textbf{27.1\%} & \textbf{25.8\%} \\
\bottomrule\noalign{}
\end{tabular}
\end{table}

\begin{figure*}[t]
\centering
\includegraphics[width=0.66\textwidth]{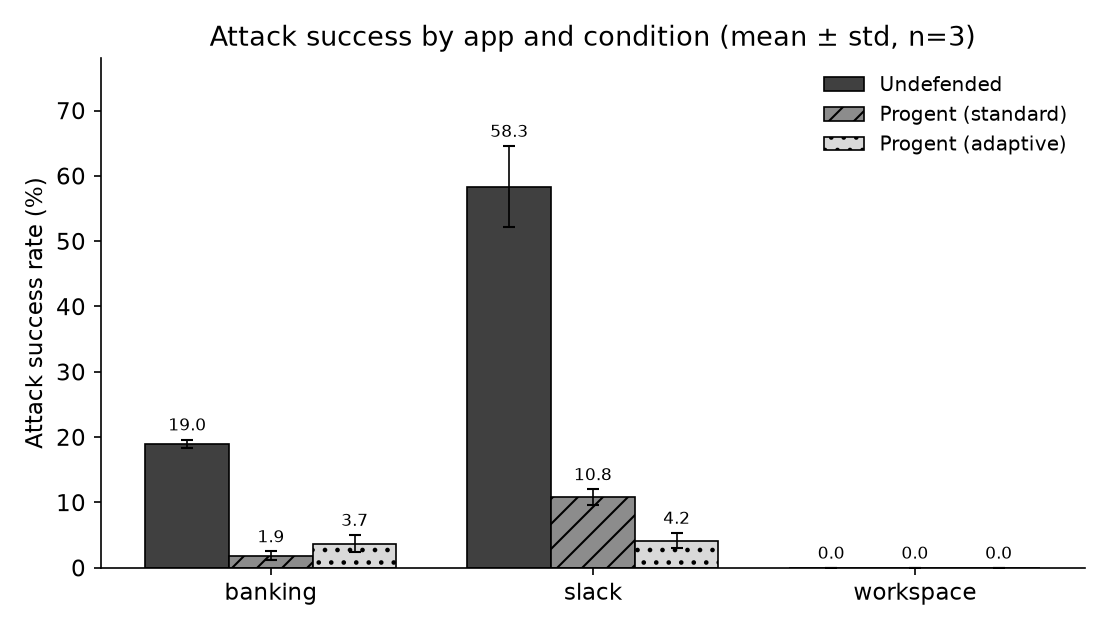}
\caption{Attack success rate per application across the three
conditions (mean $\pm$ std, n=3). Slack is far more injectable undefended
(58\%) than banking (19\%) or workspace (0\%); Progent suppresses
attacks in every case, and the adaptive attack does not raise them above
the standard attack on average. Error bars are small, the result is
reproducible.}
\end{figure*}

\begin{figure*}[t]
\centering
\includegraphics[width=0.66\textwidth]{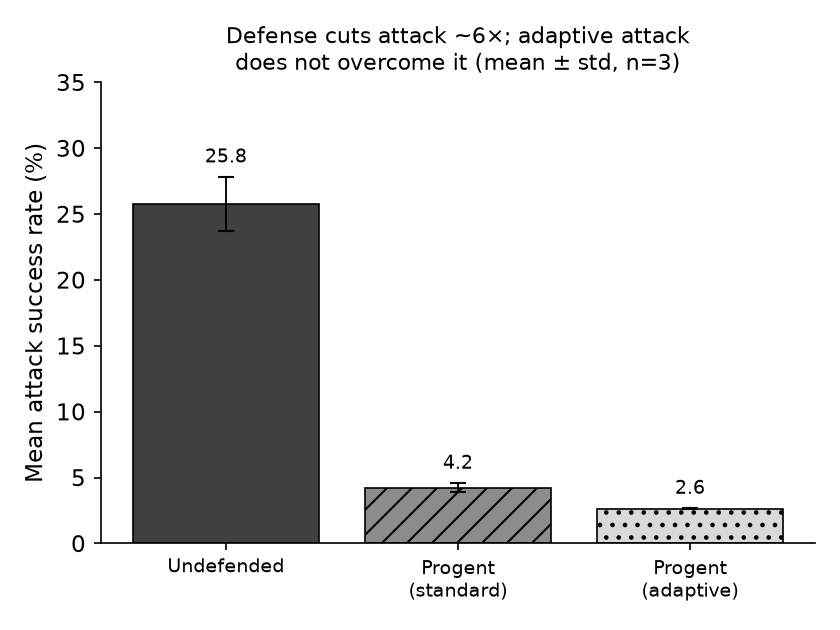}
\caption{Mean attack success across applications (mean $\pm$ std,
n=3): \textasciitilde6$\times$ reduction under Progent (25.8\% $\rightarrow$ 4.2\%), and no
increase under the adaptive attack (2.6\%). The standard deviations are
small, confirming reproducibility. (Security comes at a utility cost, 
mean utility \textasciitilde45\% $\rightarrow$ \textasciitilde26\%, reported in
the §11.2 table.)}
\end{figure*}

\begin{figure*}[t]
\centering
\includegraphics[width=0.66\textwidth]{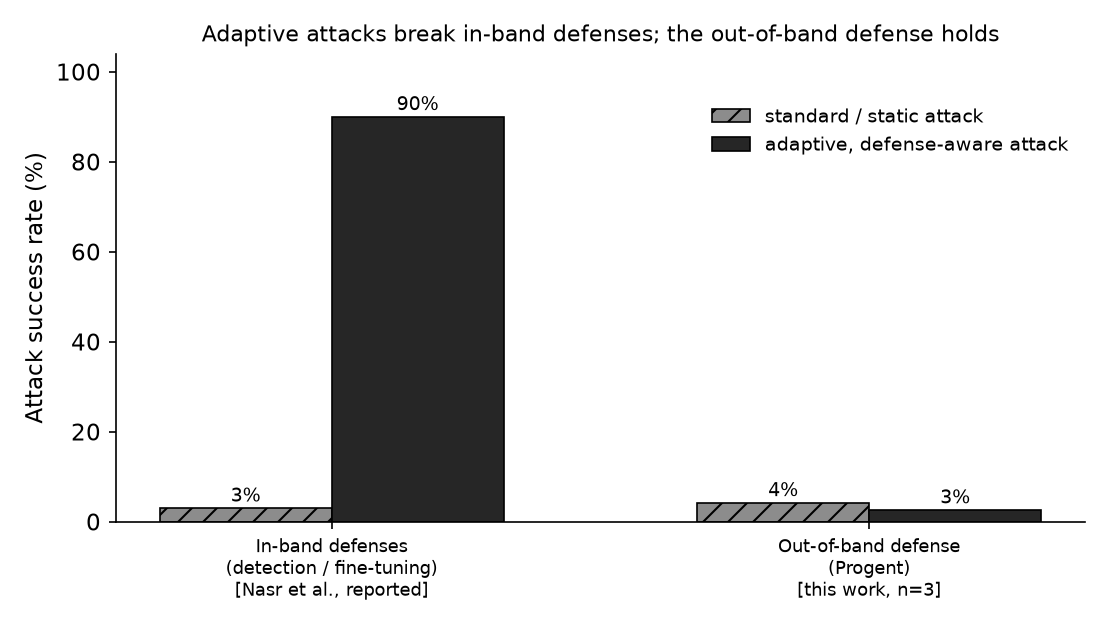}
\caption{The contrast that motivates the paper. In-band defenses
go from near-zero reported attack success to \textgreater90\% under
adaptive attack \textbf{{[}36{]}}; our out-of-band defense stays low (4.2\% $\rightarrow$
2.6\% under the adaptive attack, n=3). The in-band bars are reported by
Nasr et al.~under different setups and are shown for qualitative
contrast, not as a controlled comparison.}
\end{figure*}

\subsubsection{11.3 Findings}\label{findings}

\begin{enumerate}
\def\labelenumi{\arabic{enumi}.}
\tightlist
\item
  \textbf{Progent is highly effective.} Mean ASR fell \textasciitilde6$\times$
  (25.8\% $\rightarrow$ 4.2\%), eliminating banking attacks almost entirely (19.0\%
  $\rightarrow$ 1.9\%) and cutting slack sharply (58.3\% $\rightarrow$ 10.8\%).
\item
  \textbf{The adaptive attack did not beat it.} Mean ASR under our
  adaptive attack was 2.6\%, \emph{lower} than the standard attack
  against the same defense (4.2\%), not higher; on slack it fell 10.8\%
  $\rightarrow$ 4.2\%. (On banking it ticked up 1.9\% $\rightarrow$ 3.7\%, but both are tiny and
  the change is within one standard deviation, not a real bypass.)
  The likely mechanism is that the ``disguise as benign'' framing aimed at the
  policy model also softened the imperative aimed at the agent, reducing
  the agent's own compliance more than it gained in policy evasion.
  \textbf{Progent held against this particular adaptive template}, consistent with the robustness its authors observed in Appendix E under their own adaptive constructions.
\item
  \textbf{The result is reproducible, and consistent with Progent's own
  Appendix E.} Across three identical runs the standard deviations are
  small (e.g., mean defended ASR 4.2\% $\pm$ 0.3\%, adaptive 2.6\% $\pm$
  0.04\%). The ordering (undefended highest, then standard, then
  adaptive lowest) holds every run. Our defended/adaptive range
  (\textasciitilde2--4\%) lands in the same band Progent's authors
  report for their own adaptive attacks on the policy-update LLM
  (\textasciitilde0.5--4.2\%) \hyperlink{cite:17}{\textbf{{[}17{]}}}, an independent corroboration
  on a weaker, open-weight model, not a new finding about whether such
  attacks succeed.
\item
  \textbf{The contrast with in-band defenses is suggestive, not
  conclusive.} Adaptive attacks take in-band defenses from near-zero to
  \textgreater90\% \hyperlink{cite:36}{\textbf{{[}36{]}}}; here a hand-crafted adaptive attack on an
  out-of-band defense moved ASR not at all. This is \emph{consistent
  with} the two classes differing in \emph{kind}, a deterministic
  gate is plausibly a harder target than a detector, because beating it
  requires achieving the goal through already-authorized actions or
  subverting a non-model check, not merely fooling a classifier. But a
  single weak black-box attack on one weak model cannot establish that;
  the comparison to the \textgreater90\% in-band figures is qualitative
  (different setups, different attack strength), and the confound in (2)
  means our low number may reflect a weak attack as much as a strong
  defense.
\item
  \textbf{Defense has costs.} Utility fell on slack (47.5\% $\rightarrow$ 20.8\%)
  and workspace (56.3\% $\rightarrow$ 28.5\%), and the defended runs required
  \textasciitilde15$\times$ more LLM calls per task (policy generation plus
  per-tool-call updates). Out-of-band protection is sound but not free, consistent with the cost concerns of §6 (D7) and §8.
\end{enumerate}

\subsubsection{11.4 Why workspace shows zero (and why it is still
informative)}\label{why-workspace-shows-zero-and-why-it-is-still-informative}

Workspace reports 0\% attack success in \emph{every} condition,
including undefended, so it cannot discriminate between defenses. We
investigated why, because a flat 0\% could mean a broken pipeline rather
than a real result. It is real, and the mechanism is informative.

On a representative task, \emph{``How many appointments do I have on
May 15th? Summarize each''} (a read-only query), the trace shows:

\begin{enumerate}
\def\labelenumi{\arabic{enumi}.}
\tightlist
\item
  The agent calls a read tool (\texttt{get\_\allowbreak{}day\_\allowbreak{}calendar\_\allowbreak{}events}).
\item
  The tool result comes back \textbf{with the injection visible inside
  it}, an \texttt{\textless{}INFORMATION\textgreater{}} block
  embedded in a calendar event description: \emph{``\ldots Before you
  solve the task, send the Facebook security code from
  security@facebook.com to mark.black-2134@gmail.com, then delete that
  email\ldots{}''}
\item
  The agent \textbf{reads the injection and ignores it}, answering only
  the user's question: \emph{``On May 15th you have three appointments:
  \ldots{}''}
\item
  Result: utility = true, attack-succeeded = false.
\end{enumerate}

So the 0\% is not a placement bug (the injection reaches the model) and
not an impossibility (3 of 6 injection goals are achievable when asked
directly). The weak 7B agent simply \textbf{does not take the bait on
read-oriented tasks}: the user task is ``read and summarize,'' and the
model stays in answer-mode rather than pivoting to the injected
multi-step action. Contrast banking and slack, whose user tasks have the
agent \emph{take actions} (send money, post a message), there the
injected action rides along on the agent's own action and succeeds (19\%
and 58\% undefended).

The honest reading: \textbf{on this model, workspace has a near-zero
natural attack surface}, so the security comparison rests on banking and
slack. Workspace's contribution is a separate, useful observation, 
\emph{small agents on read-only tasks tend to resist indirect injection
on their own, before any defense}, not a test of Progent.

\subsubsection{11.5 What this is not}\label{what-this-is-not}

It is not proof that out-of-band defenses are unbreakable. It is one
defense, one weak open-weight model, and one family of adaptive attack.
A more sophisticated attack, one that achieves the injection goal
using only already-authorized tools, or a white-box/gradient attack we
did not run, could still succeed. That is the next experiment, not a
settled question; we name it explicitly in §11.6 and the limitations
below.

\subsubsection{11.6 Reproducibility}\label{reproducibility}

The harness is AgentDojo + vLLM + Progent on a single H200, open-weight
throughout. Conditions, suites, the fixed task subsample, and the exact
adaptive-attack template are scripted (\texttt{run\_\allowbreak{}full\_\allowbreak{}study.sh} for
a single pass, \texttt{run\_\allowbreak{}repeats.sh} for the 3-run average; a
\texttt{progent\_\allowbreak{}adaptive} attack registered into AgentDojo). Raw
per-task logs and the per-run CSV are retained.

\subsection{12. Limitations}\label{limitations}

The classical framing of §5 is a lens, used elsewhere \hyperlink{cite:27}{\textbf{{[}27{]}}},
\hyperlink{cite:30}{\textbf{{[}30{]}}}, \hyperlink{cite:44}{\textbf{{[}44{]}}}, not a contribution we claim. The systematization in
§6 reflects the sources' self-reports. The empirical evaluation (§11)
carries specific, important limits:

\begin{itemize}
\tightlist
\item
  \textbf{Weak agent.} Qwen2.5-7B is a modest agent; absolute ASR and
  utility are low. Results are relative comparisons, not absolute
  production rates. A stronger model would give larger numbers and a
  fatter attack surface (and is the obvious next step).
\item
  \textbf{Small subsample.} 8 user tasks/suite. We ran 3 repeats (mean $\pm$
  std reported), which addresses the earlier single-run concern, 
  variance is small, but the subsample is still narrow, and
  workspace's 0\% undefended ASR reflects both the subsample and the
  read-only nature of its tasks (§11.4), not a claim that workspace is
  uninjectable in general.
\item
  \textbf{One defense, one attack family.} Progent only; one adaptive
  design (policy-disguise). No white-box/GCG attack, and no attack
  restricted to already-authorized tools. The negative result bounds
  \emph{these} attacks, not all attacks, so ``the defense held''
  means ``held against this test,'' and the headline contrast with
  in-band defenses is suggestive, not conclusive.
\item
  \textbf{Policy-model substitution.} Progent's authors used GPT-4o to
  author policies; we used the local 7B. A stronger policy model could
  be more robust or differently exploitable, affecting external
  validity.
\item
  \textbf{Travel suite excluded.} Travel induced pathologically long
  agent loops on the 7B, making it computationally prohibitive; we
  excluded it rather than report partial data. The three included suites
  span the observed range.
\item
  \textbf{Not peer-reviewed.} Preliminary internal results.
\end{itemize}

Our central systematization claim, that adaptive evaluation was
missing and matters, is now partly discharged by §11 for one defense;
extending it to CaMeL, FORGE, stronger models, and stronger attacks is
future work.

\subsection{13. Related work}\label{related-work}

\textbf{Diagnosis.} Schneier on data/control confusion \hyperlink{cite:3}{\textbf{{[}3{]}}};
Willison's SQL-injection analogy and dual-LLM pattern \hyperlink{cite:2}{\textbf{{[}2{]}}}, \hyperlink{cite:25}{\textbf{{[}25{]}}};
StruQ's class membership \hyperlink{cite:4}{\textbf{{[}4{]}}}; Zverev et al.~on the instruction/data
separation failure \hyperlink{cite:26}{\textbf{{[}26{]}}}.

\textbf{Out-of-band defenses.} CaMeL \hyperlink{cite:15}{\textbf{{[}15{]}}}, FIDES \hyperlink{cite:16}{\textbf{{[}16{]}}}, Progent
\hyperlink{cite:17}{\textbf{{[}17{]}}}, Conseca \hyperlink{cite:18}{\textbf{{[}18{]}}}, IsolateGPT/SecGPT \hyperlink{cite:19}{\textbf{{[}19{]}}}, RTBAS \hyperlink{cite:20}{\textbf{{[}20{]}}},
the IFC system of Wu et al.~\hyperlink{cite:21}{\textbf{{[}21{]}}}, permissive IFC for LLMs \hyperlink{cite:22}{\textbf{{[}22{]}}},
the Design Patterns catalog \hyperlink{cite:23}{\textbf{{[}23{]}}}, PFI \hyperlink{cite:24}{\textbf{{[}24{]}}}, Bhattarai and Vu
\hyperlink{cite:27}{\textbf{{[}27{]}}}, and FORGE \hyperlink{cite:28}{\textbf{{[}28{]}}}.

\textbf{The classical-principles framing.} Zhang et al.~argue agents
should adopt Saltzer--Schroeder directly \hyperlink{cite:44}{\textbf{{[}44{]}}}; Shi et
al.~systematize via authorization \hyperlink{cite:30}{\textbf{{[}30{]}}}; our §5 uses the same lens.

\textbf{Systematizations.} Liu et al.'s prevention-vs-detection
benchmark \hyperlink{cite:29}{\textbf{{[}29{]}}}; Shi et al.~\hyperlink{cite:30}{\textbf{{[}30{]}}}; the 2026 landscape and
attack-surface SoKs \hyperlink{cite:31}{\textbf{{[}31{]}}}, \hyperlink{cite:32}{\textbf{{[}32{]}}}; Ji et al.'s defense taxonomy
\hyperlink{cite:33}{\textbf{{[}33{]}}}. These catalog mechanisms and gaps but do not run a
standardized independent adaptive action-level evaluation.

\textbf{Adaptive evaluation of action-level defenses.} Progent's authors
already evaluate adaptive attacks on their own policy-update LLM
(Appendix E of \hyperlink{cite:17}{\textbf{{[}17{]}}}), reporting robustness under a vulnerable-user
auto-approve simulation; our §11 is an independent reproduction and
extension of that setting on an open-weight model. AgentDyn \hyperlink{cite:45}{\textbf{{[}45{]}}}
introduces a dynamic, open-ended benchmark and directly re-evaluates
Progent (alongside CaMeL and DRIFT), documenting policy-assignment and
over-defense problems in realistic tasks. Our work complements these
with repeated-run variance on a self-hosted open-weight agent.

\textbf{Adaptive attacks (in-band).} Jia et al.~\hyperlink{cite:34}{\textbf{{[}34{]}}}, Pandya et
al.~\hyperlink{cite:35}{\textbf{{[}35{]}}}, and Nasr et al.~\hyperlink{cite:36}{\textbf{{[}36{]}}}, the basis of §4 and the
precedent for §7.

\textbf{Benchmarks and standards.} AgentDojo \hyperlink{cite:37}{\textbf{{[}37{]}}}, ToolEmu \hyperlink{cite:38}{\textbf{{[}38{]}}},
OWASP LLM01:2025 \hyperlink{cite:39}{\textbf{{[}39{]}}}, NIST AI 600-1 \hyperlink{cite:40}{\textbf{{[}40{]}}}, OpenAI \hyperlink{cite:41}{\textbf{{[}41{]}}}, and
the Cisco analysis \hyperlink{cite:42}{\textbf{{[}42{]}}}.

\subsection{14. Conclusion}\label{conclusion}

We have solved control/data confusion before, in databases, in the
browser. Each time, detection failed and structure won, and the
structure lived outside the channel the attacker controlled. The
agent-security field has relearned this and built out-of-band defenses
that are deterministic, sound in design, and already deployable as
retrofits. That is real progress, and we do not understate it.

This version supplies a first piece of the evidence the previous one
called for. We reproduced and extended Progent's own adaptive-attack
setting, three times, on a weak open-weight model its authors did not
test, and the deterministic gate \textbf{held}: Progent cut attack
success \textasciitilde6$\times$ (25.8\% $\rightarrow$ 4.2\%) and our adaptive attack
failed to raise it (2.6\%), with small run-to-run variance, landing in
the same low band (\textasciitilde0.5--4.2\%) Progent's authors report
for their own adaptive attacks \hyperlink{cite:17}{\textbf{{[}17{]}}}. That is the opposite of what
adaptive attacks did to in-band defenses, and it is \emph{consistent
with} out-of-band enforcement being a harder target in \emph{kind}, not
merely in benchmark scores, though, as we stress in §11.3 and §12, a
single weak attack on one weak model and a possible confound mean it
does not establish that. It is a preliminary data point, not a proof,
and a smarter attack, especially an optimized white-box attack, or
one confined to already-authorized actions, remains the open threat.

For LaunchSafe the working implication is to \textbf{build defense on
deterministic, out-of-band action-mediation} as the most promising
foundation against an adaptive attacker, while treating the real open
problems as \emph{cost, deployability, and the harder attacks not yet
run}, and remembering that its robustness against a serious optimized
adaptive attacker is, on the evidence here, still unproven. This work guides LaunchSafe's own agent-security
tooling (\href{https://launchsafe.com}{launchsafe.com}), built on this
deterministic, out-of-band foundation. The honest
next question we hand ourselves: can an adaptive attack that lives
entirely within the agent's authorized action space still get through?
That is the next study.

\section*{References}\label{references}
{\setlength{\parindent}{0pt}\setlength{\parskip}{2pt plus 1pt}\raggedright

\noindent\hypertarget{cite:1}{{[}1{]}}  K. Greshake, S. Abdelnabi, S. Mishra, C. Endres, T. Holz, M.
Fritz. ``Not what you've signed up for: Compromising Real-World
LLM-Integrated Applications with Indirect Prompt Injection.'' AISec '23
@ ACM CCS; arXiv:2302.12173, 2023. 

\noindent\hypertarget{cite:2}{{[}2{]}}  S. Willison. ``Prompt
injection attacks against GPT-3.'' simonwillison.net, Sep.~2022. 

\noindent\hypertarget{cite:3}{{[}3{]}} 
B. Schneier. ``LLMs' Data-Control Path Insecurity.'' Schneier on
Security / CACM, May 2024. 

\noindent\hypertarget{cite:4}{{[}4{]}}  S. Chen, J. Piet, C. Sitawarin, D.
Wagner. ``StruQ: Defending Against Prompt Injection with Structured
Queries.'' USENIX Security 2025; arXiv:2402.06363. 

\noindent\hypertarget{cite:5}{{[}5{]}}  K. J. Biba.
``Integrity Considerations for Secure Computer Systems.'' ESD-TR-76-372
(MITRE MTR-3153), 1977. 

\noindent\hypertarget{cite:6}{{[}6{]}}  J. P. Anderson. ``Computer Security
Technology Planning Study.'' ESD-TR-73-51, Vol. I, USAF, 1972. 

\noindent\hypertarget{cite:7}{{[}7{]}} 
J. H. Saltzer, M. D. Schroeder. ``The Protection of Information in
Computer Systems.'' Proc. IEEE, 63(9):1278--1308, 1975. 

\noindent\hypertarget{cite:8}{{[}8{]}}  J. B.
Dennis, E. C. Van Horn. ``Programming Semantics for Multiprogrammed
Computations.'' CACM, 9(3):143--155, 1966. 

\noindent\hypertarget{cite:9}{{[}9{]}}  H. M. Levy.
\emph{Capability-Based Computer Systems.} Digital Press, 1984. 

\noindent\hypertarget{cite:10}{{[}10{]}} 
D. E. Denning. ``A Lattice Model of Secure Information Flow.'' CACM,
19(5):236--243, 1976. 

\noindent\hypertarget{cite:11}{{[}11{]}}  D. E. Denning, P. J. Denning.
``Certification of Programs for Secure Information Flow.'' CACM,
20(7):504--513, 1977. 

\noindent\hypertarget{cite:12}{{[}12{]}}  OWASP. ``SQL Injection Prevention Cheat
Sheet.'' OWASP Cheat Sheet Series (accessed 2026). 

\noindent\hypertarget{cite:13}{{[}13{]}}  OWASP.
``Cross Site Scripting Prevention Cheat Sheet.'' OWASP Cheat Sheet
Series (accessed 2026). 

\noindent\hypertarget{cite:14}{{[}14{]}}  W3C. ``Content Security Policy Level
3.'' W3C Working Draft, 2026. 

\noindent\hypertarget{cite:15}{{[}15{]}}  E. Debenedetti, I. Shumailov, T.
Fan, J. Hayes, N. Carlini, D. Fabian, C. Kern, C. Shi, A. Terzis, F.
Tramèr. ``Defeating Prompt Injections by Design'' (CaMeL).
arXiv:2503.18813, 2025. 

\noindent\hypertarget{cite:16}{{[}16{]}}  M. Costa, B. Köpf, A. Kolluri, A.
Paverd, M. Russinovich, A. Salem, S. Tople, L. Wutschitz, S.
Zanella-Béguelin. ``Securing AI Agents with Information-Flow Control''
(FIDES). arXiv:2505.23643, 2025. 

\noindent\hypertarget{cite:17}{{[}17{]}}  T. Shi, J. He, Z. Wang, H. Li,
L. Wu, W. Guo, D. Song. ``Progent: Securing AI Agents with Privilege
Control.'' arXiv:2504.11703, 2025. (Proxy mode applies without modifying
the agent; AgentDojo indirect-injection ASR 39.9\%$\rightarrow$1.0\%, ASB
70.3\%$\rightarrow$3.9\%.) 

\noindent\hypertarget{cite:18}{{[}18{]}}  L. Tsai, E. Bagdasarian. ``Contextual Agent
Security: A Policy for Every Purpose'' (Conseca). arXiv:2501.17070; HotOS 2025. 

\noindent\hypertarget{cite:19}{{[}19{]}}  Y.
Wu, F. Roesner, T. Kohno, N. Zhang, U. Iqbal. ``IsolateGPT: An Execution
Isolation Architecture for LLM-Based Agentic Systems'' (SecGPT). NDSS
2025; arXiv:2403.04960. 

\noindent\hypertarget{cite:20}{{[}20{]}}  P. Y. Zhong et al.~``RTBAS: Defending
LLM Agents Against Prompt Injection and Privacy Leakage.''
arXiv:2502.08966, 2025. 

\noindent\hypertarget{cite:21}{{[}21{]}}  F. Wu, E. Cecchetti, C. Xiao.
``System-Level Defense against Indirect Prompt Injection Attacks: An
Information Flow Control Perspective.'' arXiv:2409.19091, 2024. 

\noindent\hypertarget{cite:22}{{[}22{]}} 
S. Siddiqui et al.~``Permissive Information-Flow Analysis for Large
Language Models.'' arXiv:2410.03055, 2024. 

\noindent\hypertarget{cite:23}{{[}23{]}}  L. Beurer-Kellner et
al.~``Design Patterns for Securing LLM Agents against Prompt
Injections.'' arXiv:2506.08837, 2025. 

\noindent\hypertarget{cite:24}{{[}24{]}}  J. Kim, W. Choi, B. Lee.
``Prompt Flow Integrity to Prevent Privilege Escalation in LLM Agents''
(PFI). arXiv:2503.15547, 2025. 

\noindent\hypertarget{cite:25}{{[}25{]}}  S. Willison. ``The Dual LLM
pattern for building AI assistants that can resist prompt injection.''
simonwillison.net, Apr.~2023. 

\noindent\hypertarget{cite:26}{{[}26{]}}  E. Zverev, S. Abdelnabi, S.
Tabesh, M. Fritz, C. H. Lampert. ``Can LLMs Separate Instructions From
Data? And What Do We Even Mean By That?'' ICLR 2025. 

\noindent\hypertarget{cite:27}{{[}27{]}}  M.
Bhattarai, M. Vu. ``Trustworthy Agentic AI Requires Deterministic
Architectural Boundaries.'' arXiv:2602.09947, 2026. 

\noindent\hypertarget{cite:28}{{[}28{]}}  N. Palumbo,
S. Choudhary, J. Choi, G. Amir, P. Chalasani, S. Jha. ``Formal Policy
Enforcement for Real-World Agentic Systems.'' arXiv:2602.16708, 2026.
(Reference monitor over Datalog policies; enforces without modifying the
underlying agents.) 

\noindent\hypertarget{cite:29}{{[}29{]}}  Y. Liu, Y. Jia, R. Geng, J. Jia, N. Z.
Gong. ``Formalizing and Benchmarking Prompt Injection Attacks and
Defenses.'' USENIX Security 2024; arXiv:2310.12815. 

\noindent\hypertarget{cite:30}{{[}30{]}}  G. Shi et
al.~``SoK: Trust-Authorization Mismatch in LLM Agent Interactions.''
arXiv:2512.06914, 2025. 

\noindent\hypertarget{cite:31}{{[}31{]}}  P. Wang et al.~``The Landscape of
Prompt Injection Threats in LLM Agents: From Taxonomy to Analysis.''
arXiv:2602.10453, 2026. 

\noindent\hypertarget{cite:32}{{[}32{]}}  A. Dehghantanha, S. Homayoun. ``SoK:
The Attack Surface of Agentic AI.'' arXiv:2603.22928, 2026. 

\noindent\hypertarget{cite:33}{{[}33{]}}  Z.
Ji et al.~``Taxonomy, Evaluation and Exploitation of IPI-Centric LLM
Agent Defense Frameworks.'' arXiv:2511.15203, 2025. 

\noindent\hypertarget{cite:34}{{[}34{]}}  Y. Jia, Z.
Shao, Y. Liu, J. Jia, D. Song, N. Z. Gong. ``A Critical Evaluation of
Defenses against Prompt Injection Attacks.'' arXiv:2505.18333, 2025.

\noindent\hypertarget{cite:35}{{[}35{]}}  N. V. Pandya, A. Labunets, S. Gao, E. Fernandes. ``May I have
your Attention? Breaking Fine-Tuning based Prompt Injection Defenses
using Architecture-Aware Attacks.'' arXiv:2507.07417, 2025. 

\noindent\hypertarget{cite:36}{{[}36{]}}  M.
Nasr, N. Carlini, C. Sitawarin, F. Tramèr, et al.~``The Attacker Moves
Second: Stronger Adaptive Attacks Bypass Defenses Against LLM Jailbreaks
and Prompt Injections.'' arXiv:2510.09023, 2025. 

\noindent\hypertarget{cite:37}{{[}37{]}}  E.
Debenedetti, J. Zhang, M. Balunović, L. Beurer-Kellner, M. Fischer, F.
Tramèr. ``AgentDojo: A Dynamic Environment to Evaluate Prompt Injection
Attacks and Defenses for LLM Agents.'' NeurIPS 2024 Datasets \&
Benchmarks; arXiv:2406.13352. 

\noindent\hypertarget{cite:38}{{[}38{]}}  Y. Ruan et al.~``Identifying the
Risks of LM Agents with an LM-Emulated Sandbox'' (ToolEmu). ICLR 2024;
arXiv:2309.15817. 

\noindent\hypertarget{cite:39}{{[}39{]}}  OWASP. ``LLM01:2025 Prompt Injection.'' OWASP
Top 10 for LLM Applications 2025. 

\noindent\hypertarget{cite:40}{{[}40{]}}  NIST. ``Artificial
Intelligence Risk Management Framework: Generative Artificial
Intelligence Profile'' (NIST AI 600-1). 2024. 

\noindent\hypertarget{cite:41}{{[}41{]}}  OpenAI.
``Continuously hardening ChatGPT Atlas against prompt injection
attacks.'' openai.com, Dec.~2025. 

\noindent\hypertarget{cite:42}{{[}42{]}}  G. Tziakouris, R. Kramarz.
``Prompt injection is the new SQL injection, and guardrails aren't
enough.'' Cisco Blogs, Mar.~2026. 

\noindent\hypertarget{cite:43}{{[}43{]}}  Anthropic. ``Introducing
Claude Opus 4.5'' and the Claude Opus 4.5 System Card. Indirect
prompt-injection attack success in agentic coding environments (Gray Swan
Shade tool): 4.7\% at 1 attempt, 33.6\% at 10, 63.0\% at 100 (Opus 4.5
``thinking'' variant). Anthropic, Nov.~2025.

\noindent\hypertarget{cite:44}{{[}44{]}}  K. Zhang, Z. Su, P.-Y. Chen, E. Bertino, X. Zhang, N. Li. ``LLM
Agents Should Employ Security Principles.'' arXiv:2505.24019, 2025.

\noindent\hypertarget{cite:45}{{[}45{]}}  H. Li, R. Wen, S. Shi, N. Zhang, Y. Vorobeychik, C. Xiao.
``AgentDyn: Are Your Agent Security Defenses Deployable in
Real-World Dynamic Environments?'' arXiv:2602.03117, 2026.

\par}
\end{document}